\documentstyle[11pt,newpasp,twoside,epsf]{article}
\def\msol{\ifmmode {\>M_\odot}\else {$M_\odot$}\fi}
\def\cmsq{\ifmmode {\>{\rm\ cm}^2}\else {cm$^2$}\fi}
\def\psqcm{\ifmmode {\>{\rm cm}^{-2}}\else {cm$^{-2}$}\fi}
\def\psqpc{\ifmmode {\>{\rm pc}^{-2}}\else {pc$^{-2}$}\fi}
\def\pcsq{\ifmmode {\>{\rm\ pc}^2}\else {pc$^2$}\fi}
\def\Tkev{\ifmmode{T_{\rm kev}}\else {$T_{\rm keV}$}\fi}
\def\hubunits{\ifmmode {\>{\rm km\ s^{-1}\ Mpc^{-1}}}\else {km
s$^{-1}$ Mpc$^{-1}$}\fi}
\def\gta{\;\lower 0.5ex\hbox{$\buildrel > \over \sim\ $}}
\def\lta{\;\lower 0.5ex\hbox{$\buildrel < \over \sim\ $}}

\def\phiIV{\ifmmode{\varphi_4}\else {$\varphi_4$}\fi}
\def\phiI{\ifmmode{\varphi_i}\else {$\varphi_i$}\fi}

\def\be{\begin{equation}}
\def\ee{\end{equation}}
\def\bea{\begin{eqnarray}}
\def\eea{\end{eqnarray}}
\def\beas{\begin{eqnarray*}}
\def\eeas{\end{eqnarray*}}
\def\gtrapprox{\;\lower 0.5ex\hbox{$\buildrel >\over \sim\ $}}
\def\lessapprox{\;\lower 0.5ex\hbox{$\buildrel < \over \sim\ $}}
\def\deg   {$^\circ$}

\def\Em    {${\cal E}_m$}

\def\tauLL {\ifmmode{\tau_{\scriptscriptstyle LL}}\else 
           {$\tau_{\scriptscriptstyle LL}$}\fi}

\def\Em{\ifmmode{{\rm E}_m}\else {{\rm E}$_m$}\fi}
\def\NH{\ifmmode{{\rm N}_{\scriptscriptstyle\rm H}}\else {{\rm N}$_{\scriptscriptstyle\rm H}$}\fi}

\def\Ha    {H$\alpha$}

\def\HI    {H${\scriptstyle\rm I}$}
\def\HII    {H${\scriptstyle\rm II}$}
\def\eg    {{\it e.g., }}
\def\ie    {{\it i.e., }}
\def\cf    {{\it cf. }}
\def\qv    {{\it q.v., }}
\def\etal  {\ et al.}
\def\kms{\ifmmode {\>{\rm km\ s}^{-1}}\else {km s$^{-1}$}\fi}

\def\Em{\ifmmode{{\cal E}_m}\else {{\cal E}$_m$}\fi}
\def\Dm{\ifmmode{{\cal D}_m}\else {{\cal D}$_m$}\fi}
\def\fesc{\ifmmode{\hat{f}_{\rm esc}}\else {$\hat{f}_{\rm esc}$}\fi}
\def\fescs{\ifmmode{f_{\rm esc}}\else {$f_{\rm esc}$}\fi}
\def\rsolar{\ifmmode{r_\odot}\else {$r_\odot$}\fi}
\def\emunit{\ifmmode{{\rm cm}^{-6}{\rm\ pc}}\else {
cm$^{-6}$ pc}\fi}
\def\intensity{\ifmmode{{\rm erg\ cm}^{-2}{\rm\ s}^{-1}
      {\rm\ Hz}^{-1}{\rm\ sr}^{-1}}
      \else {erg cm$^{-2}$ s$^{-1}$ Hz$^{-1}$ sr$^{-1}$}\fi}
\def\flux{\ifmmode{{\rm erg\ cm}^{-2}{\rm\ s}^{-1}}\else {erg
cm$^{-2}$ s$^{-1}$}\fi}
\def\fluxdensity{\ifmmode{{\rm erg\ cm^{-2}\ s^{-1}\ Hz^{-1}}}\else {erg
cm$^{-2}$ s$^{-1}$ Hz$^{-1}$}\fi}
\def\phoflux{\ifmmode{{\rm phot\ cm}^{-2}{\rm\ s}^{-1}}\else {phot
cm$^{-2}$ s$^{-1}$}\fi}
\def\phorate{\ifmmode{{\rm phot\ s}^{-1}}\else {phot s$^{-1}$}\fi}

\begin{document}
\title{\Ha\ Distance Constraints for High Velocity Clouds in the Galactic Halo} 
\author{Joss Bland-Hawthorn}
\affil{Anglo-Australian Observatory, PO Box 296, Epping, NSW 2121, Australia}
\author{Philip R. Maloney}
\affil{Center for Astrophysics \& Space Astronomy, University of Colorado, Boulder, CO 80309-0389}

\begin{abstract}
We present some developments in determining \Ha\ distances
to high-velocity clouds (HVCs) in the Galactic halo. Until recently, it
was difficult to assess the nature and origin of HVCs because so
little was known about them. But now several HVCs have reliable distance
bounds derived from the stellar absorption technique, and more than a 
dozen have abundance measurements. In addition, twenty or more HVCs have
been detected in \Ha\ (and a few in optical forbidden lines). Over
the past five years, we have been developing a model of the halo radiation
field which includes contributions from the stellar disk, the stellar
bulge, the hot corona, and the Magellanic Clouds.\footnote{The 
{\tt diskhalo} ionization code, along with full documentation 
(Bland-Hawthorn \& Maloney 2001), is to be 
made available for general use.} In certain instances, the \Ha\ flux 
from an opaque \HI\ cloud can be used to derive a crude distance 
constraint to the cloud. For a UV escape fraction of \fesc\ $\approx$ 
6\% perpendicular to the disk (\fescs\ $\approx$ $1-2$\% when averaged 
over solid angle), the HVCs appear to be broadly consistent with the 
spiral arm model. We caution that a larger database with full sky 
coverage is required before the usefulness of \Ha\ distances can be 
fully assessed. We present a number of detailed predictions from our
distance frame to encourage independent assessments from future 
observations.  If the model is valid, we find that most HVCs detected
to date are scattered throughout the halo up to distances of 50 kpc from 
the Sun.  Most of this material is likely to be debris from recent
galaxy interactions, or even debris dislodged from the outer Galaxy
disk.  We propose some future tests of the \Ha\ distance model and
briefly discuss recent \Ha\ detections along the Magellanic Bridge and
Magellanic Stream.
\end{abstract}

\section{Introduction}\label{intro}
Observations  of the  Galactic halo  make a  compelling case  that the
formation of halos continues to  the present day (Wyse 1999). The halo
appears to have built up through a process of accretion and merging of
low-mass  structures  which  is  still   going  on  at  a  low  level.
Hierarchical cold dark matter (CDM) simulations, however, predict that
the Galactic halo  should have many more satellites  than are actually
observed (Klypin et al. 1999; Moore et al.1999).

As much as 40\% of the sky is peppered with high-velocity \HI\ clouds 
(HVCs) which  do  not  conform  to  orderly  Galactic  rotation 
(Wakker \& van Woerden 1991; Putman 2000).   These  are
interesting  accretion   candidates  $-$  particularly   if  they  are
associated  with  dark  matter  `mini  halos' $-$  except  that  their
distances, $d$, are  unknown for all but a few  sources.  As a result,
fundamental  physical  quantities  $-$  size ($\propto  d$)  and  mass
($\propto  d^2$)  $-$  are  unconstrained which  has  encouraged  wide
speculation  as to the  nature of  HVCs (Wakker  \& van  Woerden 1997).

Indeed, the current  renaissance in HVC studies  can be traced in  part 
to an interesting suggestion by Blitz et al. (1999).  They showed that
the velocity centroids and
groupings  of positive/negative  velocity  clouds on  the  sky may  be
understood  within  a reference  frame  centered  on  the Local  Group
barycenter (\cf Zwaan \& Briggs 2000).  They interpret  HVCs as  gas 
clouds  accreting  onto the
Local  Group over  a megaparsec  sphere. Braun  \& Burton  (2000) 
identify  specific examples  of compact  clouds that  have `rotation
curves' consistent with CDM mass profiles. For sources at 700~kpc, the
kinematic  signatures  imply  a  high dark-to-visible  mass  ratio  of
10$-$50.

This paper addresses the distance problem. In earlier papers (\eg\ 
Bland-Hawthorn et al. 1998; Bland-Hawthorn \& Maloney 1999 $-$ BM99), we showed
that faint \Ha\ measures from distant \HI\ clouds could, in principle,
be used to estimate crude distances to the clouds. The \Ha\ emission measure 
from any cloud which can be detected at 21~cm is a direct measure of the 
Lyc (Lyman continuum) radiation field, independent of distance. In order 
to interpret the \Ha\ emission, we require a realistic model of the 
Galactic halo ionizing field. Here, we present some recent developments 
in deriving \Ha\ distances which significantly extends our earlier work.

\section{The Escape of UV radiation from the Galaxy}\label{escape}

Since we have discussed the UV escape fraction (\fesc)
from the Galaxy at length in earlier articles (\qv\ Bland-Hawthorn \& 
Putman 2001), only a short discussion is given here. 
The likely value of \fesc\ is still uncertain but there is increasing
evidence that UV does radiate far from HII regions. To cite one of
several recent demonstrations, the warm (diffuse) ionized medium in 
external galaxies clearly shows the hallmark of the HII regions even
after these have been masked out of the \Ha\ image 
(\eg\ Zurita et al. 2001; Cianci 2001). 

Further evidence that UV must escape the Galaxy comes from
the measured electron density profile from halo pulsars. Manchester \&
Taylor (2000; see also Nordgren, Cordes \& Terzian 1992) have modelled
this with a scale height of 800~pc which exceeds or is comparable to
the scale height of the diffuse \HI\ (warm neutral medium; Lockman
1984). Without fine-tuning, it is unlikely that the Reynolds layer
represents a radiation-bounded medium within a co-extensive \HI\
envelope. We know that the radiation field must be soft from the
weakness of HeI$\lambda$5876 and non-detection of [HeII]$\lambda$4686
(Reynolds \& Tufte 1995). Furthermore, the observed weakness of
[OI]$\lambda$6300/H$\alpha$ indicates two things: (i) the ionization
fraction must be high (Reynolds 1989), (ii) all of the UV photons
produced in the disk cannot be absorbed in radiation-bounded \HII\
regions (Domg\"{o}rgen \& Mathis 1994).

It is important to appreciate that the Fabry-Perot `staring' method is
so sensitive that we only need \fescs\ $\sim$ 1\% for the \Ha\ distance 
method to be useful. In fact, our preliminary analysis of \Ha\ 
levels from \HI\ clouds with stellar distance brackets suggests that indeed
\fescs\ $\approx$ 1$-$2\% (see \S~6).
At distances of 300 kpc along the polar axis, the Galaxy field is
comparable to the cosmic UV field. The expected \Ha\ levels are of
order $1-2$ mR and it may be possible to reach these 
levels\footnote{1 milliRayleigh $=$ $10^3/4\pi$ phot cm$^{-2}$ s$^{-1}$ 
sr$^{-1}$ $=$ $2.4\times 10^{-10}$ erg cm$^{-2}$ s$^{-1}$ sr$^{-1}$ at
\Ha.}
with new differential techniques (\eg\ Glazebrook \& Bland-Hawthorn 2001).

\begin{figure}\label{fig1}
\plotfiddle{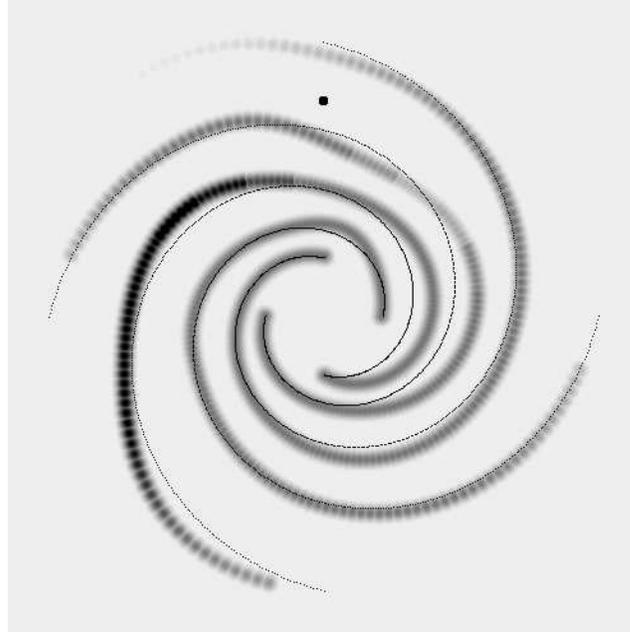}{3in}{0}{48}{48}{-150}{-70}
\caption{The spiral arm distribution of UV emitting sources
used in the {\tt diskhalo} ionization code where the dot shows the
Sun's position. The model is identical to the Taylor-Cordes density 
distribution in the near field (r$<$10 kpc). In the far field, the 
spiral arms are forced to follow the Ortiz-Lepine model shown as
continuous lines. }
\end{figure}

\begin{figure}\label{fig2}
\plotfiddle{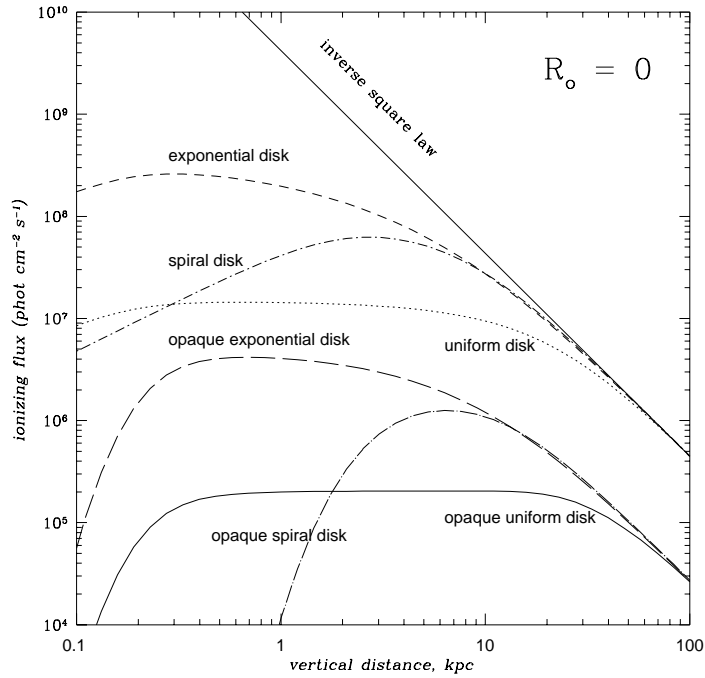}{3.1in}{0}{48}{48}{-150}{-80}
\caption{
A comparison of the halo ionizing flux for different disk distributions 
(uniform emissivity, exponential and spiral)
within the {\tt diskhalo} code compared to a simple inverse square law. 
The vertical distance is measured from the center of the disk along the 
polar axis. The top three curves are in the absence of dust and converge 
in the far field limit. The lower three curves include the effects of
dust where \tauLL $=$ 2.8.
}
\end{figure}

\section{The importance of spiral arms}\label{spiral}

Our early attempts to derive \Ha\ distances made use of a smooth 
exponential distribution of ionizing sources (BM99).
For HVC distances within 10 kpc of the plane, we need a more realistic 
distribution of ionizing sources than the exponential disk model. 
This would be straightforward if we knew the
exact location of all O stars, and the precise dust distribution
throughout the Galaxy. But this will not be possible until the GAIA
astrometric mission flies in 2010. However, most studies of spirality
in the Galaxy agree that the tangent points of the spiral arms are
well defined over a wide range of methods and wavelengths, in particular,
the distribution of pulsar dispersion measures with galactic longitude
(Taylor \& Cordes 1993).

For this reason, the non-axisymmetric component of the {\tt diskhalo} 
ionization model links its fortunes to the standard
model for determining pulsar distances. Rough distances to pulsars are
deduced from the dispersion (and scattering) measure due
to warm electrons along the line of sight.  Early attempts used a smooth
distribution of electrons (\eg Manchester \& Taylor 1981) although
Lyne, Manchester \& Taylor (1985) showed that typical distance estimates
have random errors as large as a factor of two.
After the inclusion of smooth spiral arms, Taylor \& Cordes (1993) predict
that most distances should be good to $\sim$20\%.  This level of accuracy
is somewhat surprising when one examines face-on spirals in the
Ultraviolet Imaging Telescope (UIT) database. But the distance model is 
largely borne out by lower limits derived from pulsar sight lines which 
show \HI\ in absorption.

\section{Stellar bulge contribution}

A component largely overlooked to date is the UV field arising from
the Galactic spheroid (O'Connell 1999). A `UV upturn' component
is now well established for
ellipticals and S0s (Greggio \& Renzini 1999; Macchetto et al. 1996;
Bica et al. 1996). Binette et al. (1994) suggest this may arise from post-AGB
stars evolving from an old stellar population.  A rough estimate is
$7\times 10^{50} (M_{\rm bulge}/10^{10} M_\odot)$ ionizing phot s$^{-1}$.
$M_{\rm bulge}$ is the mass of the Galactic halo bulge which is uncertain
to a factor of two, \ie\ $M_{\rm bulge} \approx 1-2 \times 10^{10} M_{\odot}$
(Dwek et al. 1995; Zhao 1996).  Since much of the bulge stars lie outside most
of the absorbing ISM, the UV flux which escapes into the halo may be
non-negligible and more smoothly distributed compared to flux which
escapes the disk. When a bulge component is discussed below, 
we include an isothermal distribution of sources with a total luminosity
of $7\times 10^{50}$ \phorate. We assume that all of the UV escapes,
although the dusty disk blocks half of the radiation.  The hot corona 
component is discussed in BM99.  Both sources of UV are examined briefly 
in the next section.

\section{The disk-halo ionization model}\label{model}

The {\tt diskhalo} model includes five basic components:
the spiral arm (or exponential) disk defined by
the OB star population; stellar bulge; hot coronal halo; LMC (and
SMC); cosmic background. In addition, the resulting radiation field
can be moderated by the presence of a plane parallel opacity law,
and projected \Ha\ emission measures can be derived with a variety
of schemes.

In Fig.~1, we show how the spiral arms are represented in the ionization
model. In the near field (r $<$ 10 kpc), the distribution is 
essentially identical to the Taylor-Cordes model. However, the spiral
arms are incomplete on the far side of the Galaxy which required us
to extend the arm coverage by 60\%.  This was done by splicing the 
Taylor-Cordes model onto the basic model parameters of Ortiz \& Lepine 
(1993). The spiral arms cover a total area of 100 kpc$^2$ within a 
circle with 12 kpc radius, compared with the exponential model which 
extends out to a radius of 25 kpc (although most of the emissivity 
lies within a 12 kpc radius).

The {\tt diskhalo} manual 
(Bland-Hawthorn \& Maloney 2001) describes how the electron density field is 
inverted to produce the surface density of ionizing photons.  Fig.~2 shows 
how the three different disk distributions compare as we approach the 
disk along the polar axis of the Galaxy.  Fig.~3 presents a cross section
through the halo field and shows that the form of the dusty spiral is
very different to that of the exponential disk within 10 kpc, which can
also be seen in Fig.~2.

An important prediction of the spiral arm and the exponential models
is that ionizing flux seen by a cloud `inside' the Solar Circle should
be much higher than the flux seen by a cloud on the `outside'. We have
tried to illustrate this effect in Fig.~4. For this trend to be evident,
a large nearby sample of clouds is required over a wide solid angle. One
possible population which might reveal this effect are the intermediate
velocity clouds (IVCs).  In the next section, we show the prediction
for one such cloud, Complex K.

Fig.~4 shows how a stellar UV bulge (or, for that matter, a hot UV
corona) can wash out some of the contrast produced by spiral arms
(see also Fig.~8).
The presence of a UV-bright stellar bulge can remove the 
near-field solution which typically arises with the spiral arm models.
This phenomenon is seen in the exponential disk predictions in Fig.~5.
The hot galactic corona is not as effective as the stellar bulge
at washing out the spiral structure since, even though it has a 
scale length a factor of 5 larger than the bulge, the expected flux 
levels are lower by about the same factor.

\begin{figure}\label{fig3}
\plotfiddle{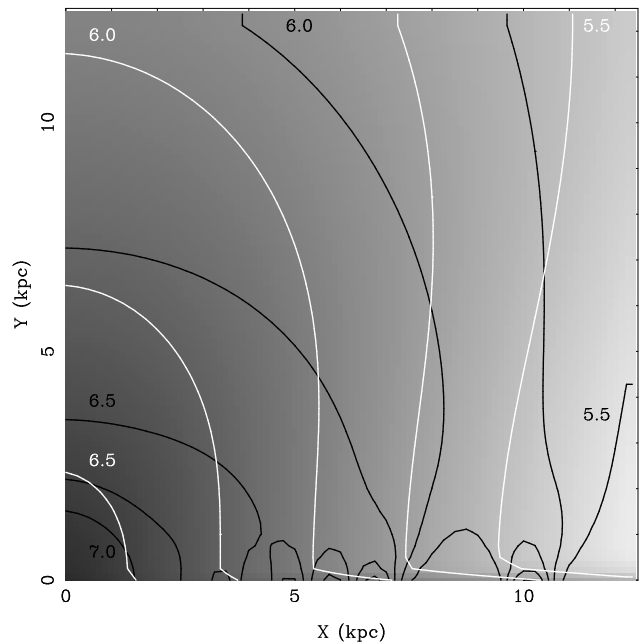}{3.1in}{-90}{48}{48}{-180}{270}
\caption{
A cross section through the halo field produced by the {\tt diskhalo} 
model for two cases: the dusty exponential disk (white contours and halftone) 
and the dusty spiral disk (black contours). The Galactic Center is at (0,0). 
The numbers give the ionizing flux in units of log(\phoflux).
}
\end{figure}

\begin{figure}\label{fig4}
\plotfiddle{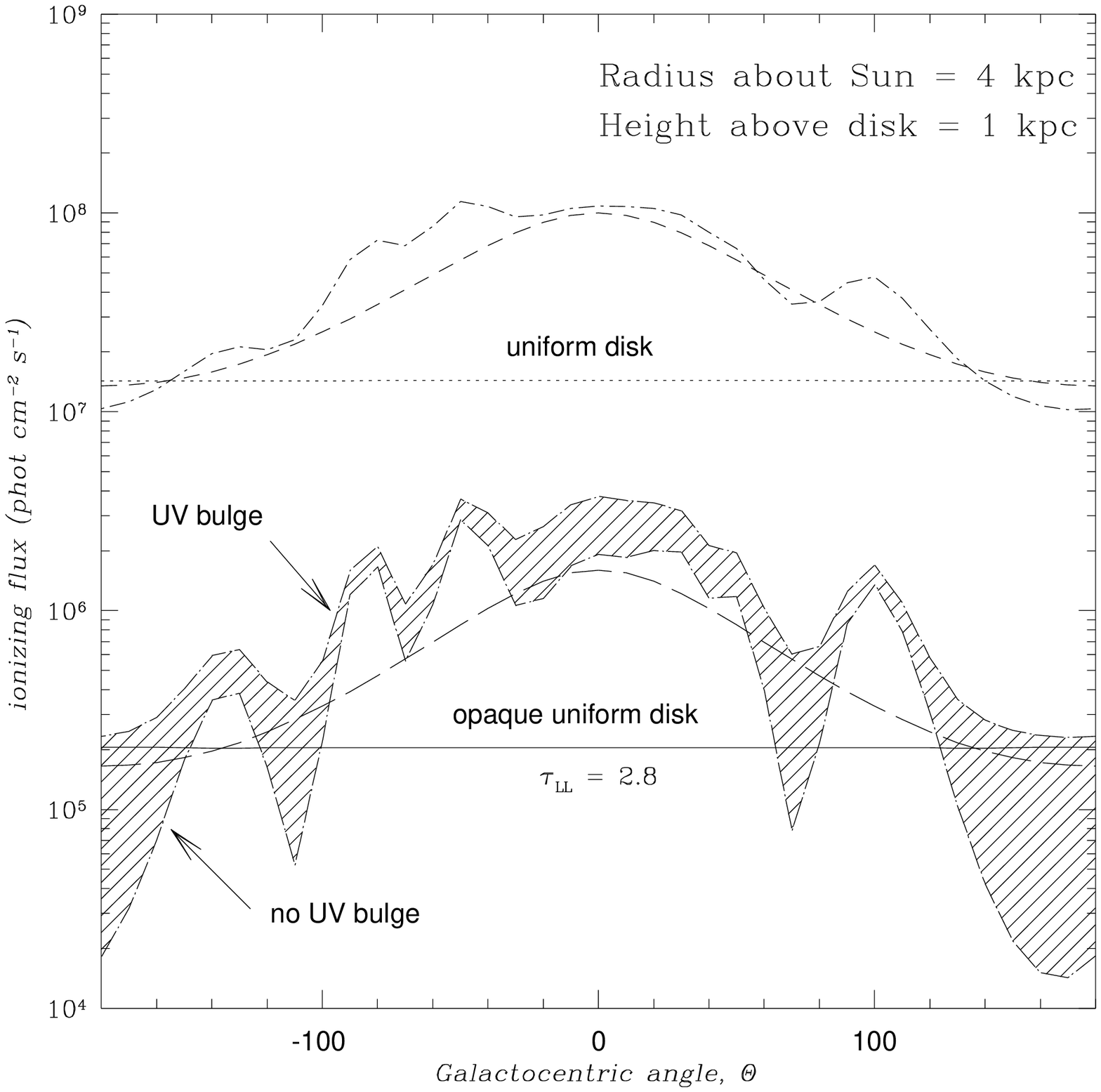}{3.1in}{0}{48}{48}{-145}{-80}
\caption{
The predicted ionizing flux around a ring at 1 kpc elevation with 
a radius of 4 kpc centered on the Sun ($\Theta=0$ is towards the 
Galactic Center). The upper curves show the
flux for optically thin cases (\tauLL $=$ 0) of a uniform, exponential
and spiral disk; the lower curves are for the dusty models (\tauLL $=$
2.8) of the same. The dusty spiral disk has been computed with and 
without the presence of a stellar UV bulge. The expected flux levels
are much higher in the direction of the Galactic Center (`inside')
compared to the Anti-Center (`outside').
}
\end{figure}

\section{Predicted \Ha\ distances to \HI\ clouds}\label{predict}

Here we present predictions for HVCs with measured \Ha\ fluxes.
We emphasize that the method is only intended as a statistical
constraint. Even with large numbers of \Ha\ detections for each
cloud complex,  it may only be useful to within a factor of a few
and even this level of confidence needs to be tested.

We used the {\tt diskhalo} code with the dusty spiral disk
(\tauLL $=$2.8). We use a conservative model with a hot corona
but no stellar bulge component.  The choice of \tauLL\ is discussed 
in detail elsewhere but largely arises from the WHAM detections
of Complexes A, C, K and M (see below). Weiner\etal\ (2001, these
proceedings) 
derive a similar value from an analysis of their own 
observations.  Since galactic coordinates have not yet been 
published for some of the detections,
we have averaged our calculations over the observed \HI.

A major benefit of the spiral arm model is the order
of magnitude larger contrast in \Ha\ that one obtains compared to the 
exponential disk model. A good example of this comes from the Complex L
sight line presented in Fig.~5. This cloud is found to be very bright
in \Ha\ by both the Las Campanas (Weiner\etal\ 2001) and TAURUS teams 
(Putman\etal\ 2001, in prep.). The exponential model fails by a large factor to 
explain the signal whereas this is a natural consequence of the spiral
arm model if the complex is 1 kpc from the disk, in which case
the cloud system lies directly above a spiral arm. Independent support
for this comes from the observation by both teams that the [NII]/\Ha\ ratios
are very elevated in this system, an effect seen at low latitudes in
external edge-on galaxies. The same temperature effect (Reynolds,
Haffner \& Tufte 1999) is seen in the Smith Cloud which is also 
thought to lie within a few kpc of the disk (see below).

But the contrast effect comes at a price. For most of the modelled
sight lines in Fig.~5, there is a near-field and a far-field 
prediction.
This is because in almost any direction away from the Sun, the radius 
vector crosses a spiral arm. Multiple solutions are generally avoided 
at latitudes higher than 20\deg\ because as we move away from the disk,
the details of the distribution become less important. An example
is the Smith Cloud (GCP) discussed in Bland-Hawthorn\etal\ (1998).
This cloud is predicted to lie at 1 kpc or roughly 15 kpc. Two
of the sight lines (Complex L, Co-Rotate) are multiple valued for
the observed \Ha.

The WHAM detections of Complexes A, M and C are also modelled in Fig.~5.
These clouds are particularly important as they have distance bounds
from the stellar absorption line (SAL) method. If a star of known distance
lies beyond the cloud, and another star of known distance falls in front,
we obtain a distance bracket for the cloud if it is seen in absorption
against the distant star, but not in the nearer star. 
Complex M (specifically, Cloud MII) has an upper limit of 4 kpc on its 
distance (Ryans\etal\ 1997); no lower bound exists. The far-field limit
is only consistent at the 1.3$\sigma$ level.
Complex A lies between 4 and 10 kpc from the Sun (van Woerden\etal\ 
1999) which appears consistent with the far field limit. There is a 
suggestion that this range could be tightened to 8 and 10 kpc
should the complex not be detected on the spectrum of PG 0832+675,
in which case the far field prediction is a factor of two too small.
Complex C has a secure lower limit on its distance of 1 kpc,
based upon five stellar probes; a much weaker limit
of $>$6~kpc is provided by the non-detection of the cloud in 
BS 16034-0114 (Wakker 2001). Both near- and far-field solutions are 
consistent with a firm distance limit ($>$1 kpc); only the far-field 
solution is consistent with the weaker limit ($>$6 kpc).

The WHAM team have recently published \Ha\ observations for the
IVC Complex K (Haffner\etal\ 2001). There is little absorption line
data in this direction although the cloud does appear to lie between
0.3 and 7.7 kpc.  The sight line to Complex K lies directly over the 
tangent point of a spiral arm which explains the broad distribution 
in Fig.~5. The WHAM team present three averaged \Ha\ measurements for 
this cloud. In the absence of a characteristic \Ha\ 
measurement for a fully mapped cloud, we take the peak \Ha\ detection 
towards the center of a cloud as the most important for distance 
determination.\footnote{This emphasizes a weakness in the \Ha\ method $-$
see Bland-Hawthorn \& Maloney 1999$b$ for a detailed discussion $-$ in that a 
detailed mapping is normally required to reliably interpret what part 
of the cloud is being lit up in \Ha. However, this is not always a 
problem. Weiner\etal\ (2001, these proceedings) and Tufte\etal\ (1998) 
find that the 
internal \Ha\ dispersion in a significant number of HVCs can be small.}
The brighter detections are entirely consistent with the SAL bounds,
although the weaker detections would push the far-field limit out to
10 kpc or more.

\medskip
In summary, for HVCs with well defined distance bounds, we find 
(see also Weiner\etal\ 2001) that the observed H$\alpha$ is
roughly consistent with $\hat{f}_{\rm esc}$ $\approx$ 6\% (\tauLL $=$ 2.8)
although the present uncertainties are about a factor of 2. Note that
this escape value
$\hat{f}_{\rm esc}$ is defined orthogonally to the disk plane;
the solid-angle averaged value is $f_{\rm esc}$ $\approx$ 1$-$2\%.

Finally, the Las Campanas team (Weiner et al. 2001) have achieved some 
faint detections to clouds which are thought to be farther afield. 
We show the predictions for three 
of these (Co-Rotate, GCN and Population N). We have run the code for
all HVCs with \Ha\ detections to date. Population N appears to lie at
the greatest distance (30$-$45 kpc) if the \Ha\ emission arises from
the disk radiation field.

\begin{figure}
\plotfiddle{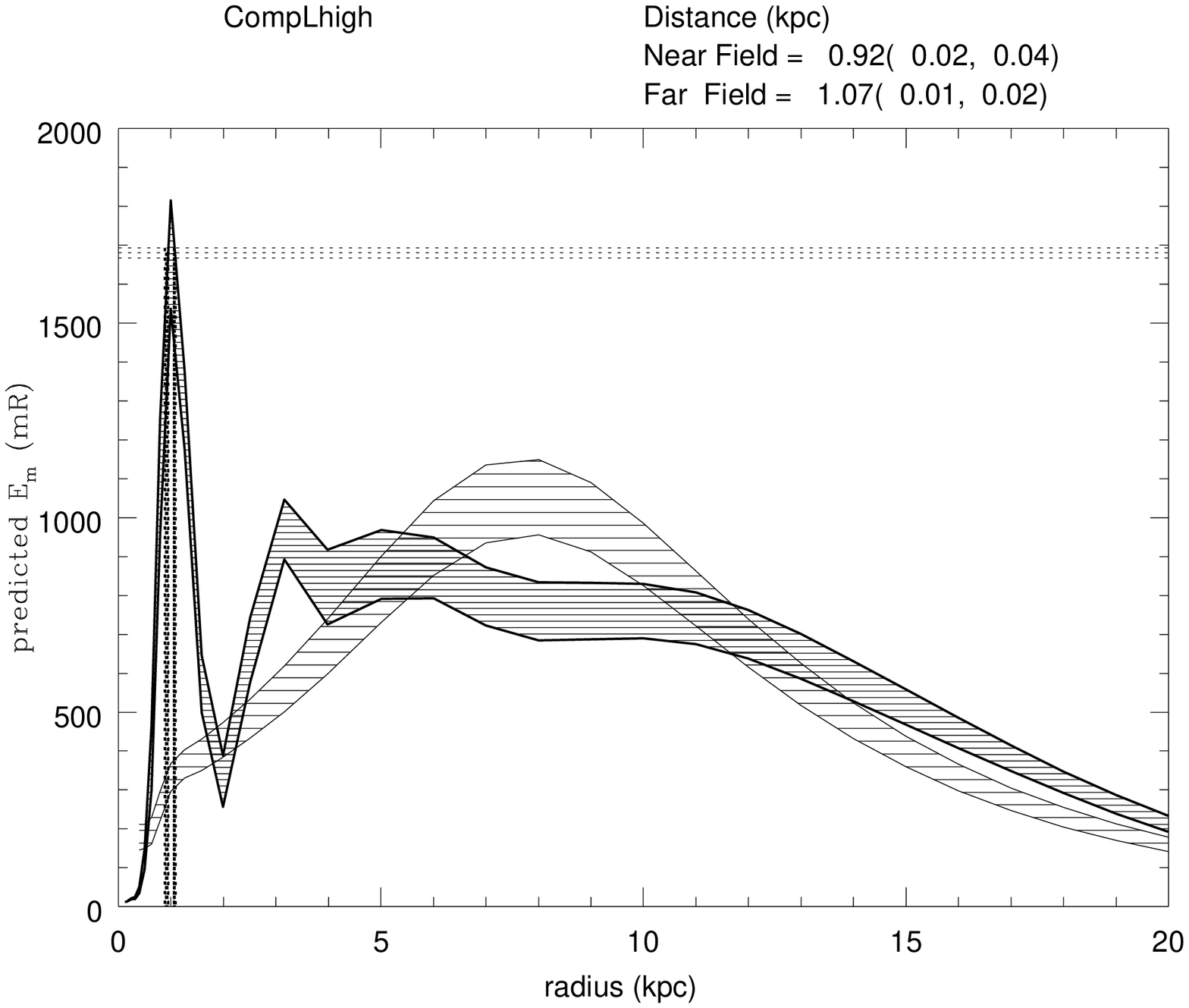}{0.64in}{0}{25}{25}{-155}{-85}
\plotfiddle{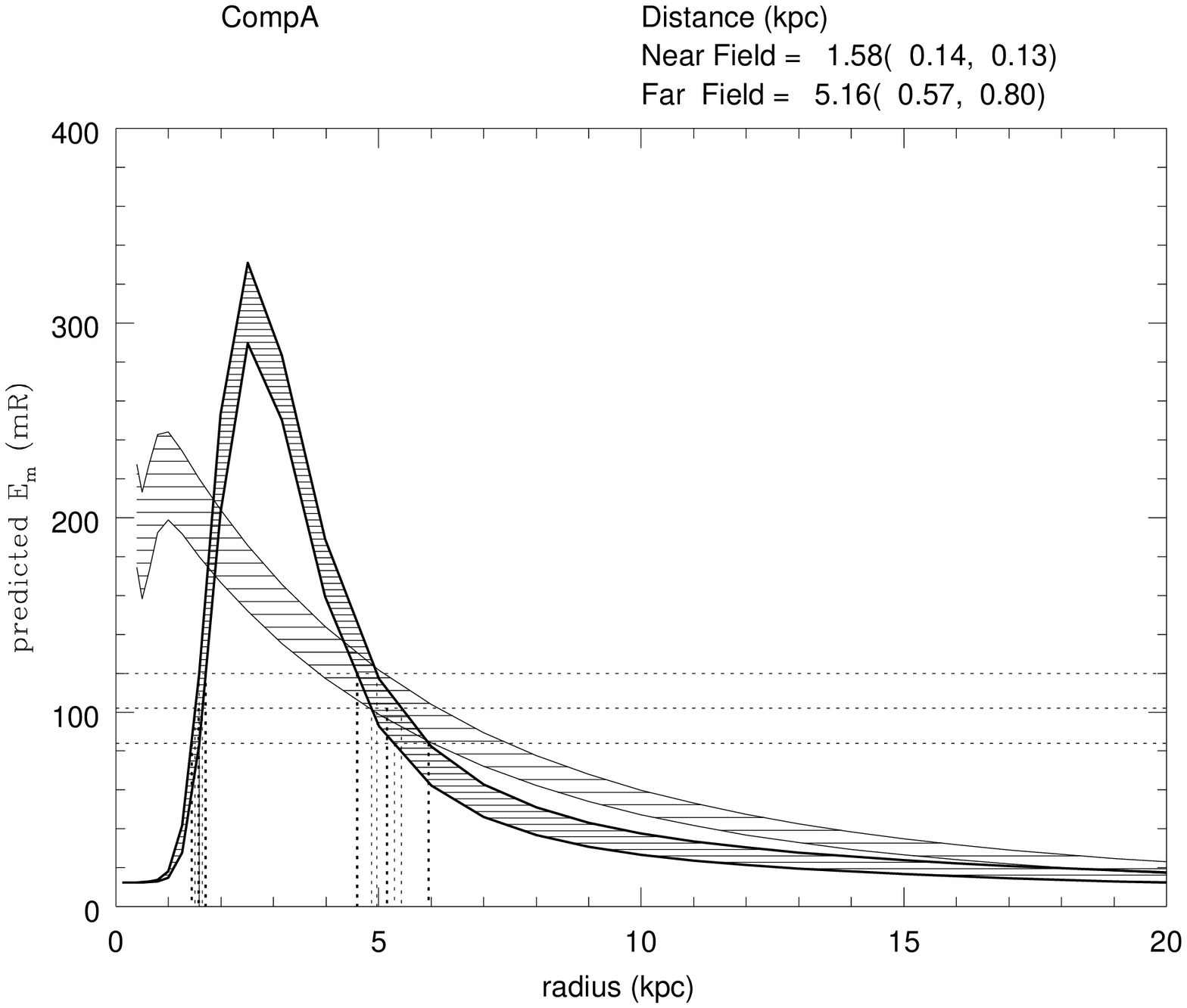}{0.64in}{0}{25}{25}{-5}{-25}
\plotfiddle{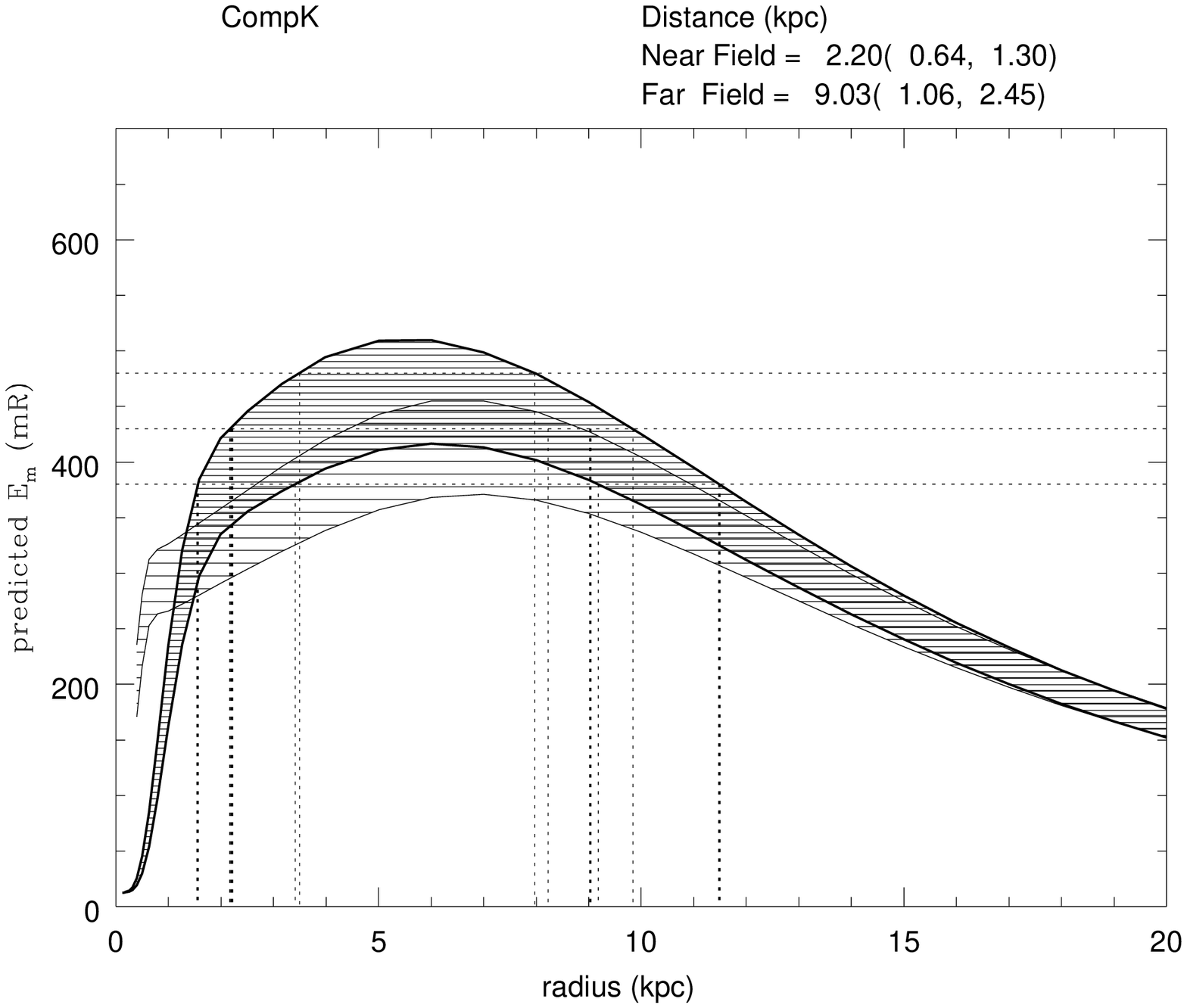}{0.64in}{0}{25}{25}{-155}{-85}
\plotfiddle{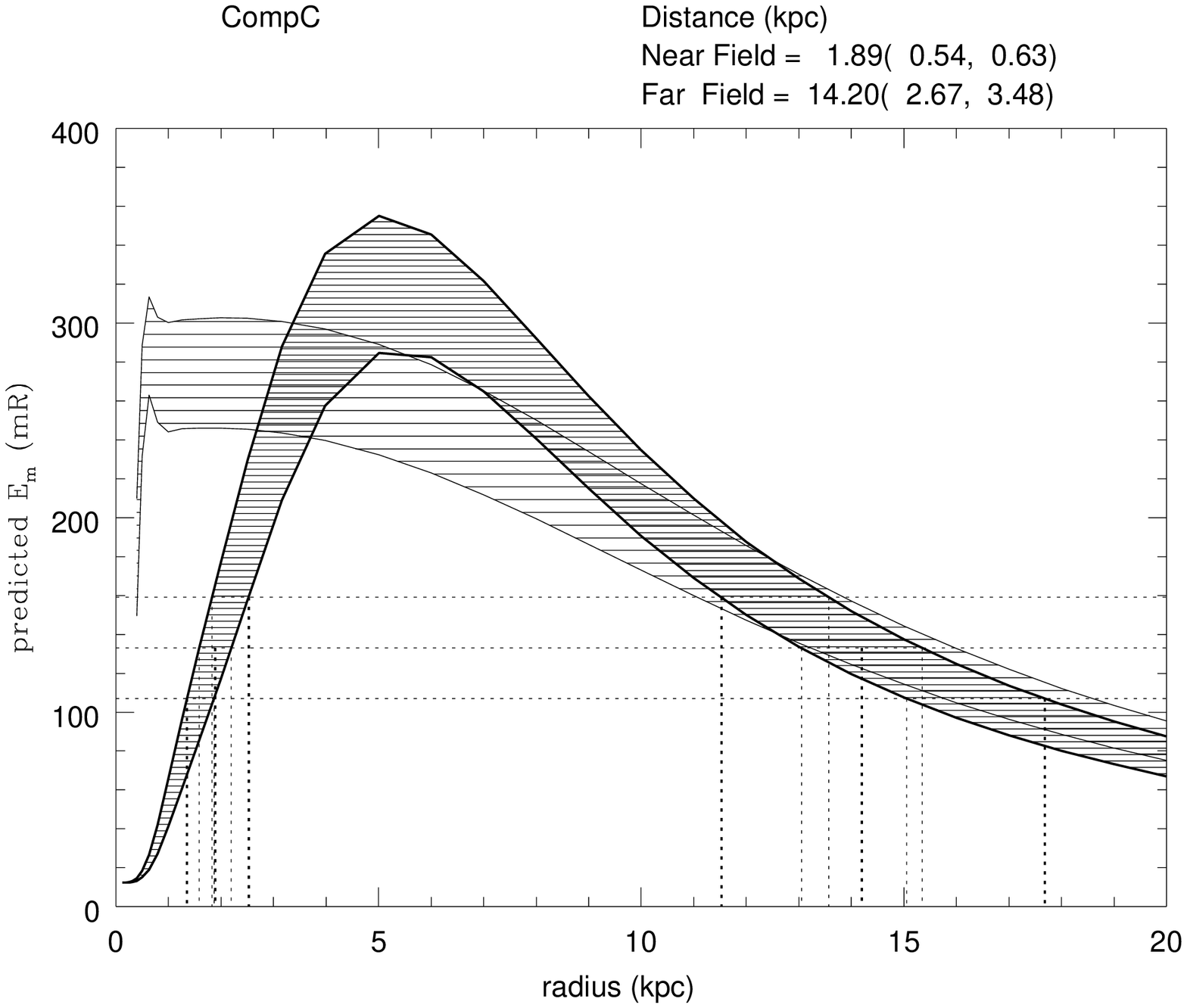}{0.64in}{0}{25}{25}{-5}{-25}
\plotfiddle{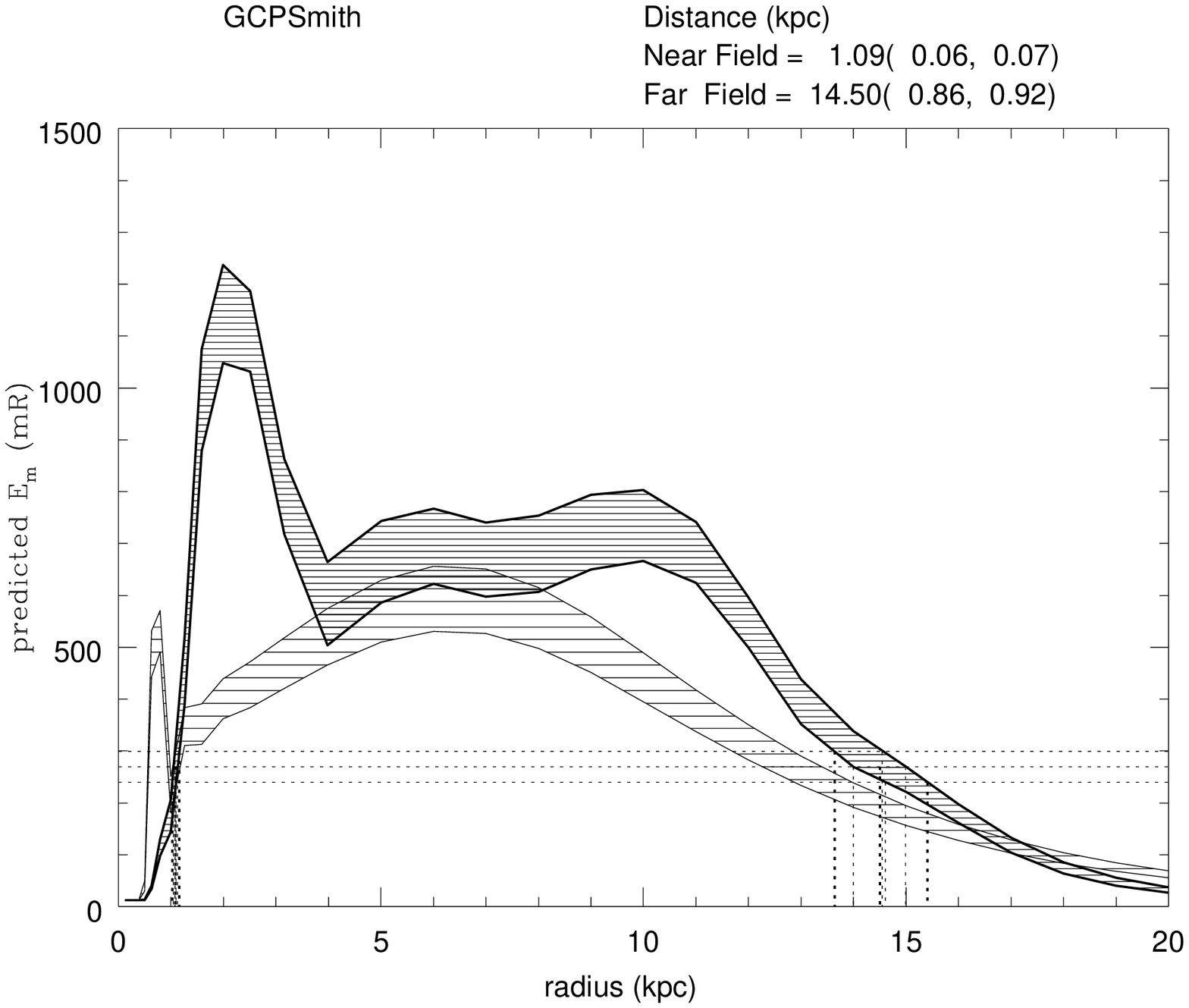}{0.64in}{0}{25}{25}{-155}{-85}
\plotfiddle{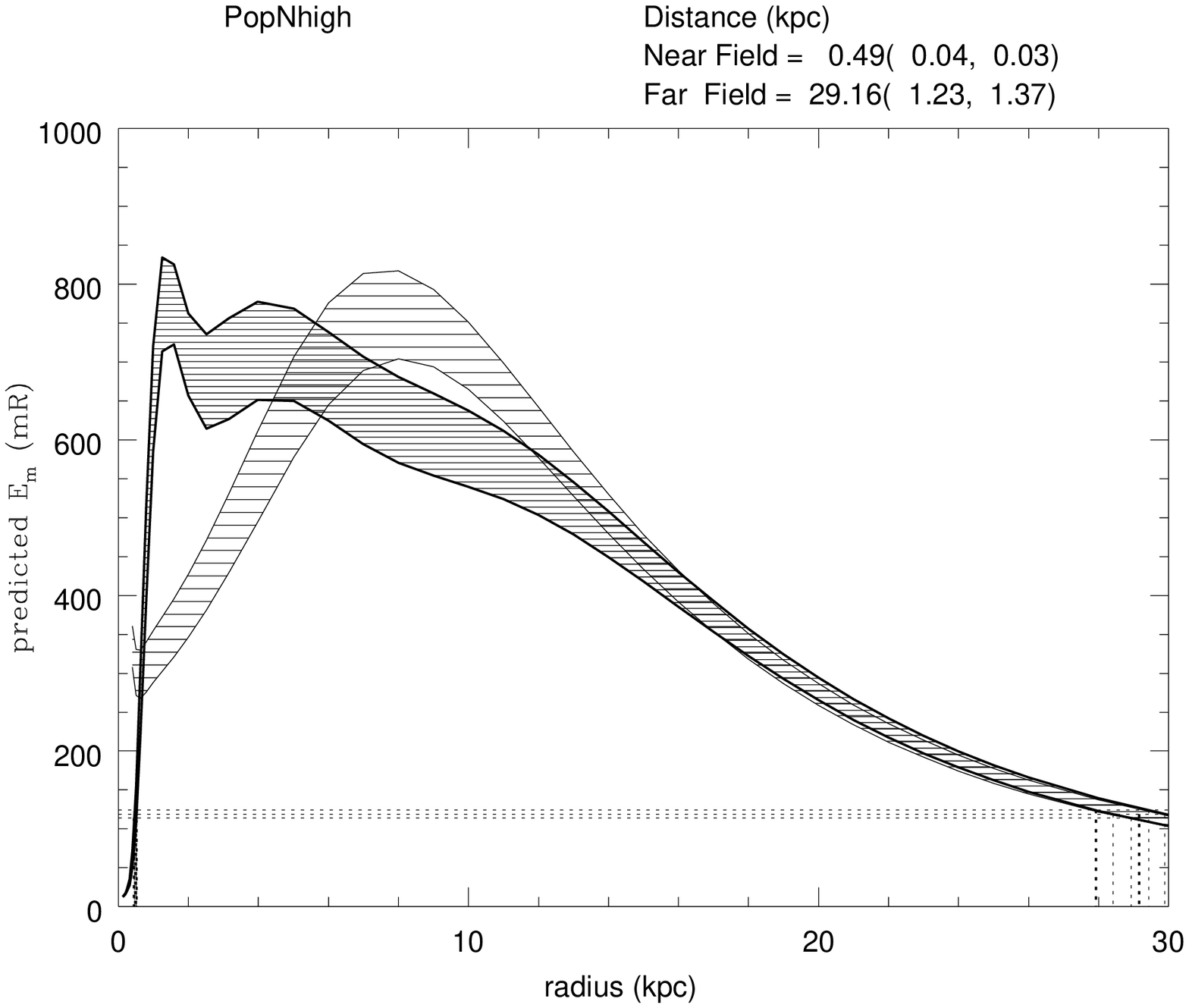}{0.64in}{0}{25}{25}{-5}{-25}
\plotfiddle{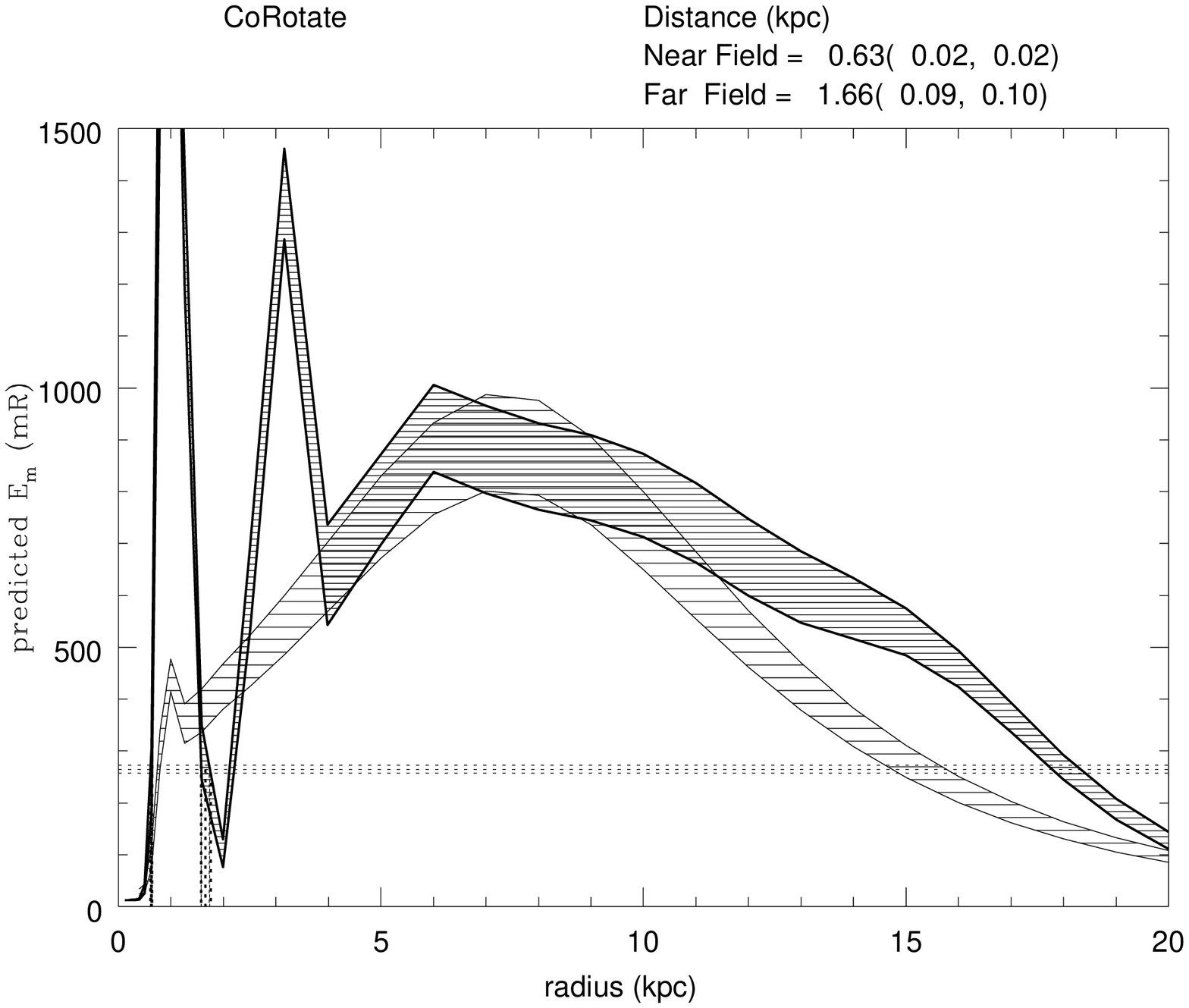}{0.64in}{0}{25}{25}{-155}{-85}
\plotfiddle{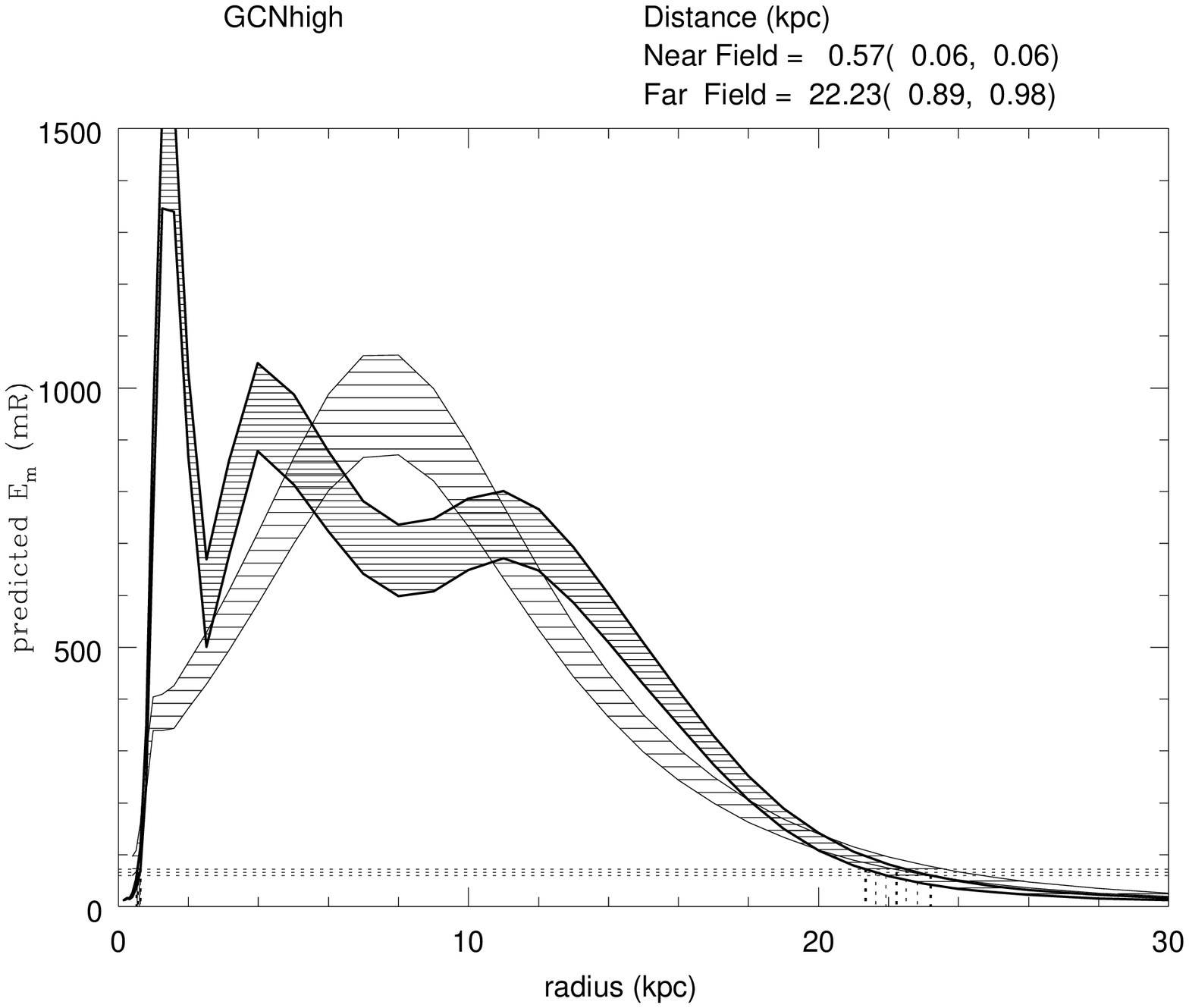}{0.64in}{0}{25}{25}{-5}{-25}
\caption{Predicted emission measures along the radius vector to 8 HVCs. The
densely shaded band arises from the dusty spiral disk model averaged over
different cloud orientations; the lighter shaded band is for a dusty
exponential disk. The horizontal lines show the average and range of 
emission measures for the modelled position on the cloud. The vertical 
lines show where the horizontal lines intersect the model. For the spiral
model, there is almost always a near-field and a far-field solution
although low latitude vectors can produce multiple solutions.
}
\end{figure}

\section{The Magellanic System}\label{magellan}

The LMC has several highly active star forming regions,
particularly regions of very recent star formation (Shapley III) and
of ongoing star formation (30 Doradus).  
The basic ionizing requirement of the LMC from combined UV, optical and
radio studies appears to be $5\times 10^{51}\,\phoflux$. Within a factor
of two, this is consistent with OB star counts (Walborn 1984; Parker 1993),
radio continuum observations (McGee, Brooks \& Batchelor 1972;
Israel \& Koornneef 1979), and vacuum ultraviolet observations (Smith et al.
1987) of the LMC. However, the total number of ionizing photons
produced by the LMC \HII\ regions, spread over a 5 kpc region,
may be as high as $1.5-3\times 10^{52}\,\phorate$ (Smith et al. 1987).
OB star counts around 30 Dor (\eg Parker 1993) could well underestimate the
total ionizing flux by a substantial factor.  Kennicutt et al. (1995) suggest 
that fully one third of the ionizing radiation in the LMC arises from within 
0.5\deg\ of 30 Dor.  The ground-based results may suffer from crowding 
which means that the total number of stars is underestimated.

We now test whether the Magellanic \HI\ Bridge can constrain
the escape fraction of UV photons from the Magellanic Clouds. Fig.~6
shows one such model prediction for the Bridge. Fujimoto \& Sofue (1976) 
give specific positions for the LMC and SMC in Galactic coordinates.
Here, we take the total
ionizing flux from the LMC to be $1\times 10^{52}\;\phorate$ and the SMC
to be an order of magnitude smaller. For both galaxies, we assume that
15\% of the UV escapes. The expected level in the Bridge is of order 
$50-100$ mR, an order of magnitude higher than levels produced by the 
Galaxy at that large polar angle.

There are two published claims of \Ha\ in the Magellanic Bridge.
Johnson, Meaburn \& Osman (1982) claim to see diffuse \Ha\ across the
entire HI bridge extending over 30\deg\ or more at an observed surface
brightness of 8~R$\pm$4~R.  Marcelin, Boulesteix \& Georgelin (1985) 
detected the Shapley wing of the SMC in \Ha\ at the level of 4~R.
These \Ha\ detections, which are partly confirmed by the UKST \Ha\ 
survey (Zealey 2001, personal communication), are more than an order of 
magnitude stronger than expected in our simple model proposed. It seems
that some of the bright \Ha\ regions arise from embedded UV sources,
\eg\ UV-bright stars in the Shapley wing. However, this does not appear
to be the explanation for recent detections of bright \Ha\ along
the Magellanic Stream (Weiner \& Williams 1996; Weiner et al. 2001, these
proceedings).

Since the Magellanic Stream passes directly over the South Galactic Pole
(illustrated in Fig.~7), BM99 had hoped this would constrain \fesc\ from 
the Galaxy but their original model underestimates the required flux by at 
least a factor of 5. Weiner et al. (2001) claim detections to some
points of the Stream as high as 1~R. The Galaxy fails to produce this level
of ionization by at least an order of magnitude.  The expected \Ha\ levels 
produced by the Galaxy are shown in Fig.~8 where we also show the additional 
contribution from the LMC.  Since the LMC and much of the Stream lie close to 
$Y=0$, the model predictions are given in this plane.  Bland-Hawthorn \&
Putman (2001) discuss various dynamical scenarios which may account for the 
bright Stream detections.

\bigskip
In summary, the Magellanic Stream poses a clear problem for the \Ha\ distance 
method since, in a few locations, it appears to be an order of magnitude 
brighter than predicted by the model. The Magellanic Bridge is also \Ha\ 
bright in places, but this may be due to internal UV sources. However, 
the HVCs appear to be broadly consistent with the spiral arm ionization 
model, and we find that most HVCs detected to date are scattered throughout
the halo up to distances of 50 kpc from the Sun.  Our suspicion is that
most of the HVCs, like the Magellanic Bridge and the Magellanic Stream, 
result from galaxy interactions, specifically dwarf galaxies disrupted
by the Galaxy. An interesting prospect that we have been investigating 
with B.K. Gibson is that some of the HVCs have been dislodged from the 
outer disk due to the disruptive passage of a nearby dwarf. There are
several observations which support this: (i) the  metallicities are 
consistent with the outer disk; (ii) the outer HI disk looks highly
disturbed (Burton \& Te Lintel Hekkert 1986); (iii) in some instances 
the HVCs show continuity in velocity with the outer disk 
(\eg\ Gibson et al. 2001); and (iv) the \Ha\ distances place many of
the clouds on 10 kpc scales.  We have begun to carry out hydrodynamical 
simulations in order to test this idea further.

\begin{figure}
\plotfiddle{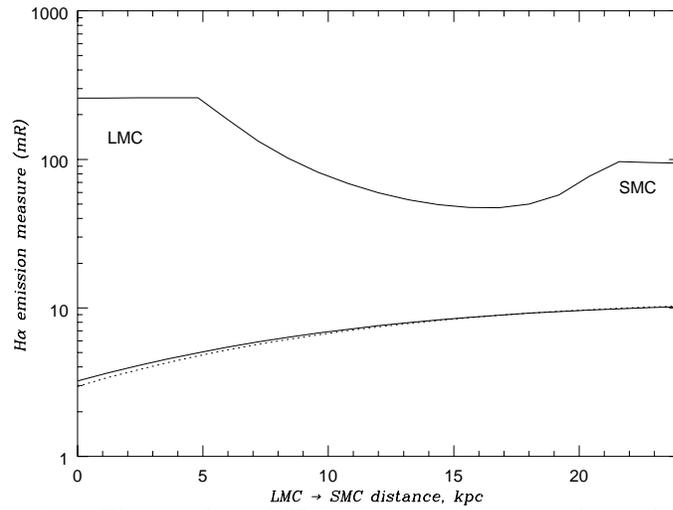}{3in}{0}{48}{48}{-150}{-88}
\caption{The predicted \Ha\ emission measure
along the Magellanic Bridge in units of log(mR). The LMC 
is to the left and the SMC is to the right.  The upper curve is due
to escaping radiation from the Magellanic Clouds.
The lower curves are due the dusty exponential and spiral
disks (\tauLL $=$ 2.8).\label{fig6}
}
\end{figure}

\begin{figure}
\plotfiddle{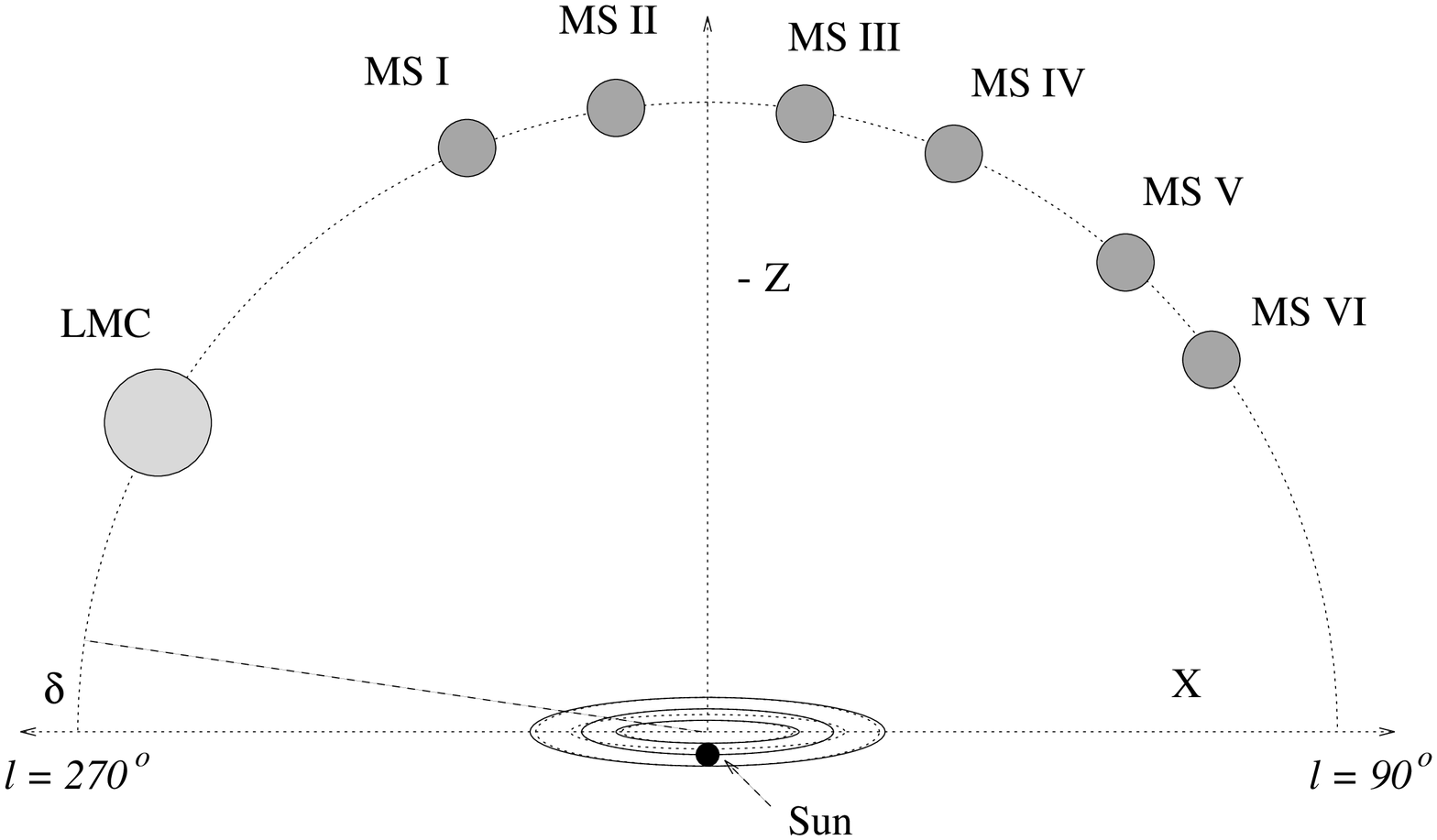}{3.1in}{0}{48}{48}{-150}{0}
\caption{
An illustration of the LMC and the dominant clouds in
the Magellanic Stream (Mathewson \& Ford 1984) projected onto the Galactic
$X$-$Z$ plane. The orbit of the Stream lies closer to the Great Circle whose
longitude is $l = 285^\circ$.  We ignore small projection errors resulting
from our vantage point at the Solar Circle. The angle $\delta$ is measured
from the negative $X$ axis towards the negative $Z$ axis where
$\delta = -b\ (0^\circ \leq \delta \leq 90^\circ)$ and
$\delta = b+180^\circ\ (90^\circ \leq \delta \leq 180^\circ)$.  
\label{fig7}
}
\end{figure}
\begin{figure}\label{fig8}
\plotfiddle{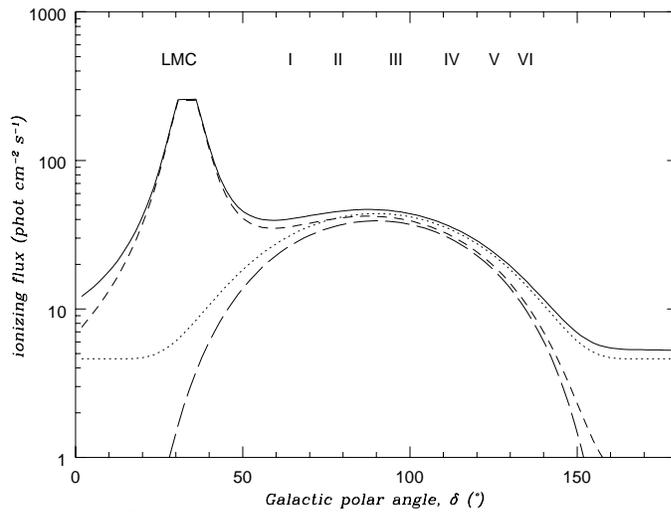}{3.1in}{0}{48}{48}{-150}{-88}
\caption{
The predicted \Ha\ emission measure along the Stream
as a function of $\delta$ (Fig.~7) in units of log(mR). The 
cloud positions are illustrated in Fig.~7. All curves show
the ionizing influence of the Galaxy ($\tauLL = 2.8$). The short dashed
curve includes the contribution of the LMC; the dotted curve includes
the contribution of a UV-bright stellar bulge. The solid curve includes 
the effect of the LMC and a stellar bulge.
}
\end{figure}

\end{document}